\documentclass{ws-ijmpcs}

\begin{document}

\markboth{Simonetta Liuti}
{PQCD ANALYSIS OF PARTON-HADRON DUALITY}

%
\catchline{}{}{}{}{}
%

\title{{\bf PQCD ANALYSIS OF PARTON-HADRON DUALITY}}

\author{SIMONETTA LIUTI }

\address{{\it Department of Physics, University of Virginia, 382 McCormick Rd. \\
Charlottesville, VA 22904, USA
\\
sl4y@virginia.edu}}

\maketitle


\begin{abstract}
We propose an extraction of the running coupling constant of QCD in the infrared region 
from experimental data on deep inelastic inclusive scattering at Bjorken $x \rightarrow 1$.  
We first attempt a perturbative fit of the data that extends NLO PQCD evolution to large $x$ values
and final state invariant mass, $W$, in the resonance region.
We include both target mass corrections and large $x$ resummation effects. 
These effects are of order ${\cal O} (1/Q^2)$, and they improve the 
agreement with the $Q^2$ dependence of the data. Standard analyses 
require the presence of additional power corrections, or dynamical higher twists, 
to achieve a fully quantitative fit. 
Our analysis, however, is regulated by the value of the strong coupling  in the 
infrared region that enters through large $x$ resummation effects, and that can suppress, or 
absorb, higher twist effects. Large $x$ data therefore indirectly provide a measurement of this quantity
that can be compared to extractions from other observables.

\end{abstract}

\ccode{PACS numbers: 25.30.Dh, 25.30.F,12.38.Bx, 12.38.Cyh}

\section{Introduction}	
The main goal of QCD is to describe the structure 
of hadrons in terms of its fundamental degrees of freedom given by the quarks and gluons,  or partons. Hadrons are observable  
in both the initial and final stages of a hard process 
while the existence and properties of partons are inferred only indirectly. 
Because of the smallness of the running coupling constant of QCD at large enough momentum transfer squared, $Q^2$, 
or equivalently at short distances ${\cal O}(1/\sqrt{Q^2})$, 
a hard probe sees hadrons as composed of "free" quarks and gluons carrying fractions 
$x$ of the hadron's Light Cone (LC) momentum, with given probability distributions. 
Factorization theorems regulate this fundamental property of the theory by allowing for
the short distance  behavior at a given scale, $Q^2$,
to be evaluated separately from the long distance contribution which is identified with the Parton Distribution Functions  (PDFs). 
  \cite{CTEQ_Handbook}.  
As increasingly shorter distances are probed the PDFs shape in $x$ changes due to radiative processes,  according to a pattern which is 
calculable with high accuracy within Perturbative QCD (PQCD) \cite{Moch}. 
%
 
Although the perturbative phases of a hard collision are clearly distinct from the non-perturbative stage that gives rise to hadron structure, experimental observations 
indicate  that such non-perturbative stage appears to be almost ubiquitously pre-determined by -- or to be keeping track of -- the perturbative mechanism.  
This concept goes under the name of {\it parton-hadron duality}. 
Parton-hadron duality implies in the most extreme case, that hadronic 
observables are replaced by calculable partonic ones with
little more going into the hadronic formation phase 
of the process. 
In this respect, duality can be seen as a phenomenological manifestation of the non-perturbative to perturbative transition 
in QCD. A full explanation for its onset is however still lacking.

As an example in this contribution we will focus on Bloom Gilman (BG) duality \cite{BG}, observed in DIS. 
Here the relevant kinematical variables are: $x=Q^2/2M\nu$ ($M$ being the proton mass and $\nu$ 
the energy transfer in the lab system), the four-momentum transfer, $Q^2$, and the invariant mass for the proton, $P$, and virtual photon, $q$, system, $W^2=(P+q)^2$  
($W^2 = Q^2(1/x-1)+M^2$).  For
large values of Bjorken $x  \geq 0.5$, and $Q^2$ in the multi-GeV region, one has 
$W^2 \leq 5$ GeV$^2$, {\it i.e.}  the cross section is dominated by resonance formation. While it is impossible to reconstruct the detailed structure of the
proton's resonances, these remarkably follow the PQCD predictions when averaged over  $x$ (see {\it e.g.} \cite{Mel} for a review).
Although BG duality was observed at the inception of QCD, quantitative analyses could be attempted only more recently, having at disposal the extensive, high precision data from Jefferson Lab \cite{Mel}.
Such analyses fall into two main categories:

\vspace{0.3cm}
\noindent {\it i)} Quark model/symmetries-based, Refs.\cite{CloIsg,CloMel,Jes,CloZhao}. 
These analyze the leading order matrix elements for the quark proton scattering subprocess and study the conditions under which the DIS cross section which is proportional 
to the incoherent sum of the squares of the constituent charges, can  in average connect smoothly to the resonance production portion of the cross section  
which is given by the square of the coherent sum of the constituent charges. This connection is regulated by Parity transformations. It was, in fact, shown how the excitation of resonance states of opposite Parity interferes destructively in all cases except for the leading twist. No account of perturbative processes enters this model.

\noindent {\it ii)} PQCD-based, Refs. \cite{Kep1,Kep2,BFL}.
This approach claims that only after a  complete perturbative QCD analysis  is performed can one {\em define} duality by  quantitatively establishing
whether, and to what extent, this phenomenon is responsible for the apparent cancellation of multiparton correlations.  
It was noticed that the extension of NLO PQCD evolution to large $x$ 
must include large $x$ resummation effects. The overall effect of these is to shift the scale at which $\alpha_S$ is calculated
to lower values, with increasing $x$. 
Standard analyses require the presence of additional power corrections, or dynamical higher twists, 
to achieve a fully quantitative fit. Our analysis, however, being regulated by the value of the QCD running coupling  in the 
infrared region that enters through large $x$ resummation effects, brings to a suppression of higher twist effects.
These get, in fact, absorbed in the coupling's infrared behavior.

\vspace{0.3cm}
Although these two approaches have been considered so far complementary to each other, a unified description might derive through
the definition of the effective coupling \cite{ChenDeurLiu}.  

This paper is organized as follows. In Section 2 we  present an overview of the large $x$ data, and discuss a few aspects of the evolution mechanism for 
DIS at large $x$ where two scales related to the invariant mass
and to the four-momentum transfer, are simultaneously present. We then illustrate in Section 3 the connection between large $x$ data 
and the coupling in the infrared region: we argue that once the range of validity of parton-hadron duality is defined quantitatively, 
a precise PQCD analysis at  large $x$ would open up the possibility of extracting 
the strong coupling constant at low scale. In Section 4 we draw our conclusions

\section{Analysis of large $x$ data}
Our studies of the physical origin of duality, and of its impact on our understanding of the nucleon's structure started a few years ago when we 
set up a program to quantitatively extract the scale dependence of the large $x$ Jlab Hall C data \cite{Ioana}. In the analysis performed in Refs.\cite{Kep1,Kep2} 
we addressed several effects that have a large impact at large $x$, namely Target Mass Corrections (TMCs), large $x$ resummation effects, and 
higher twists.
This work  was then completed, and extended to polarized data in Ref.  \cite{BFL}. 
An important point emerged from Refs. \cite{Kep1,Kep2,BFL} that a deeper understanding was needed 
of those aspects unique to the large $x$ perturbative QCD analysis. 

High precision inclusive unpolarized electron-nucleon scattering 
data on both hydrogen and deuterium targets from Jefferson Lab are available 
to date in the large $x$, multi-GeV regime (see \cite{Christy} and references therein).
Because of the precision of the data one should now be able to distinguish among  different sources of 
scaling violations affecting the structure functions in addition to standard NLO evolution, 
\begin{itemize}
\item Target Mass Corrections
(TMC), 
\item Large $x$ Resummation Effects (LxR)
\item Nuclear Effects 
\item Dynamical Higher Twists (HTs), 
\item Impact of Next-to-Next-to-Leading-Order (NNLO) perturbative
evolution. 
\end{itemize}
All of the effects above can be extracted with an associated theoretical error. It is in fact well known that their evaluation is model dependent.
Recent studies, however, have been directed at determining more precisely both the origin and size of the associated theoretical error. 
Recent analyses have been taking into account, so far, some but not all of the effects listed above \cite{Acc_10}. 

\subsection{Unpolarized structure function}
The inclusive DIS cross section of unpolarized electrons off
an unpolarized proton is written in terms of the two structure
functions $F_2$ and $F_1$, 
\begin{eqnarray}
\label{xsect}
\frac{d^2\sigma}{dx dy} =
\frac{4\pi\alpha^2}{Q^2 xy}
\left[
    \left(1-y-\frac{(Mxy)^2}{Q^2}\right)F_2 +
    y^2 x F_1    \right], 
\end{eqnarray}
with $y=\nu/\epsilon_1$, $\epsilon_1$ being the initial electron energy.
The structure functions are related by the equation  
\begin{equation}
\label{R}
F_1 = F_2(1+\gamma^2)/(2x(1+R)),
\end{equation}
where $\gamma^2=4M^2x^2/Q^2$, and $R$ is ratio of the longitudinal to transverse
virtual photo-absorption cross sections. 
%
In QCD, $F_2$ is expanded in series of inverse powers of $Q^2$, 
obtained by ordering the matrix 
elements in the DIS process by increasing twist $\tau$, which is equal
to their dimension minus spin
\begin{eqnarray}
\label{t-exp} 
F_{2}(x,Q^2)  =  F_{2}^{LT}(x,Q^2) +
\frac{H(x)}{Q^2} + {\cal O}\left(\frac{1}{Q^4} \right) \simeq F_{2}^{LT}(x,Q^2) \left(
1+ \frac{C(x)}{Q^2} \right) + {\cal O}\left(\frac{1}{Q^4} \right) 
\end{eqnarray}
The first term is the leading twist (LT), with $\tau=2$.
The terms of order $1/Q^{\tau-2}$, $\tau \geq 4$, in Eq.(\ref{t-exp}) 
are the higher order terms, generally referred to as 
higher twists \cite{CTEQ_Handbook}. 

\centerline{\it Target Mass Corrections} 
\vspace{0.5cm}

TMCs are included in $F_2^{LT}$.
For $Q^2$ $\geq$ 1 GeV$^2$, TMCs can be taken into
account through the following expansion \cite{DGP} 
\begin{eqnarray}
\label{TMC}
F_{2}^{LT(TMC)}(x,Q^2)  = 
    \frac{x^2}{\xi^2\gamma^3}F_2^{\mathrm{\infty}}(\xi,Q^2) + 
    6\frac{x^3M^2}{Q^2\gamma^4}\int_\xi^1\frac{d \xi'}{{\xi'}^2} 
F_2^{\mathrm{\infty}}(\xi',Q^2),
\end{eqnarray}
where $F_2^{\infty}$
is the structure function in the absence of TMCs. A more recent analysis \cite{AccQiu}
re-examined TMCs within the collinear factorization approach of \cite{ColRogSta} in order to address the 
longstanding question of  
the unphysical behavior in the threshold region of Eq.(\ref{TMC}). This originates 
from the fact that as $x \rightarrow 1$, one  
obtains a $Q^2$-dependent threshold, namely $F_2(\xi,Q^2) =0$ for  $\xi > \xi_{max} = 2/1+\sqrt{1+ 4 M^2/Q^2}$, therefore 
rendering $F_2$ undefinable as $Q^2$ varies (see discussion in \cite{Schien}).
In the formalism of Ref. \cite{AccQiu}, TMCs, applied to the helicity dependent structure functions read
\begin{equation}
F_{T}^{LT(TMC)}(x,Q^2) = \int^{\frac{1-x}{x}Q^2}_{m_\pi^2}  d m_J^2 \rho(m_J^2) F_T^\infty\left[ \xi \left(1+\frac{m_J^2}{Q^2}\right),Q^2 \right], 
\end{equation} 
where $F_T \equiv F_1$. The final quark is assumed to hadronize into a jet of mass $m_J$ with a  a process dependent distribution/smearing function $\rho(m_J^2)$.
In our extraction we take the perspective that the evaluation of TMCs is always associated with  
the evaluation of HTs --
TMCs should in principle be applied also to HTs --  in an inseparable way. Therefore  
we consistently keep terms of ${\cal O}(1/Q^4)$ \cite{BFL,AKL}, whether in the formalism/prescription of Ref. \cite{DGP} or of Ref. \cite{AccQiu}.
$H(x,Q^2)$, then, represents the ``genuine'' HT correction that involves
interactions between the struck parton and the spectators or, formally,
multi-parton correlation functions.   
%

\vspace{0.5cm}
\centerline{\it Threshold Resummation} 
\vspace{0.5cm}

In order to understand the nature of the remaining $Q^2$ dependence that cannot
be described by NLO pQCD evolution, we also include the effect of  
threshold resummation, or Large $x$ Resummation (LxR).
LxR effects arise formally from terms containing powers of 
$\ln (1-z)$, $z$ being the longitudinal 
variable in the evolution equations, that are present in 
the Wilson coefficient functions $C(z)$. 
Below we write schematically how the latter relate the parton distributions to {\it e.g.} 
the structure function $F_2$, 
\begin{equation}
F_2^{LT}(x,Q^2)  = \frac{\alpha_s}{2\pi} \sum_q \int_x^1 dz \, C(z) \, q(x/z,Q^2), 
\label{lxr}
\end{equation}   
where we have considered only the non-singlet (NS) contribution to $F_2$ since 
only valence quarks distributions are relevant in our kinematics. 
The logarithmic terms in $C(z)$ become very large at large $x$, and they need to be 
resummed to all orders in $\alpha_S$. 
Resummation was first introduced by  
linking this issue to the definition of the correct kinematical variable that determines the 
phase space for the radiation of gluons
at large $x$. This was found to be $\widetilde{W}^2 = Q^2(1-z)/z$, 
instead of $Q^2$ \cite{BroLep,Ama}. 
As a result, the argument of the strong coupling constant becomes $z$-dependent: 
$\alpha_S(Q^2) \rightarrow \alpha_S(Q^2 (1-z)/z)$ \cite{Rob,Rob1}. 
In this procedure, however, an ambiguity is introduced, related to the need of continuing 
the value of $\alpha_S$  
for low values of its argument, {\it i.e.} for $z \rightarrow 1$ \cite{PenRos}. 
Although on one side, the size of this ambiguity could be of the same order of the HT corrections 
and, therefore,  a source of theoretical error, on the other by performing an accurate analysis such as 
the one proposed here, one can extract $\alpha_S$ for values of the scale in the infrared region. 
We address this point in more detail in the next Section.

\vspace{0.5cm}
\centerline{\it Nuclear Effects} 
\vspace{0.5cm}
Theoretical uncertainties in the deuteron are taken routinely into account, and are expected
to be in sufficient control (see \cite{KulPet} and references therein).
Uncertainties arise mainly from  

\noindent 
{\it i)} Different models
of the so called nuclear EMC effect; 

\noindent 
{\it ii)} Different
deuteron wave functions derived from currently available NN potentials,
giving rise to different amounts of high momentum components;

\noindent
{\it iii)} The interplay between nucleon off-shellness and TMC in nuclei.

\vspace{0.3cm}
Finally, we did not consider NNLO calculations, these are not expected
to alter substantially our extraction since,
differently from what seen originally in the case of $F_3$, these have been proven 
to give a relatively small contribution to $F_2$.

\vspace{0.3cm}  
Once all of the above effects have been subtracted from the data, and assuming the validity 
of the twist expansion, 
Eq.(\ref{t-exp}) in this region, one can 
interpret more reliably any remaining discrepancy in terms of HTs. 
\begin{figure}
  \includegraphics[height=.35\textheight]{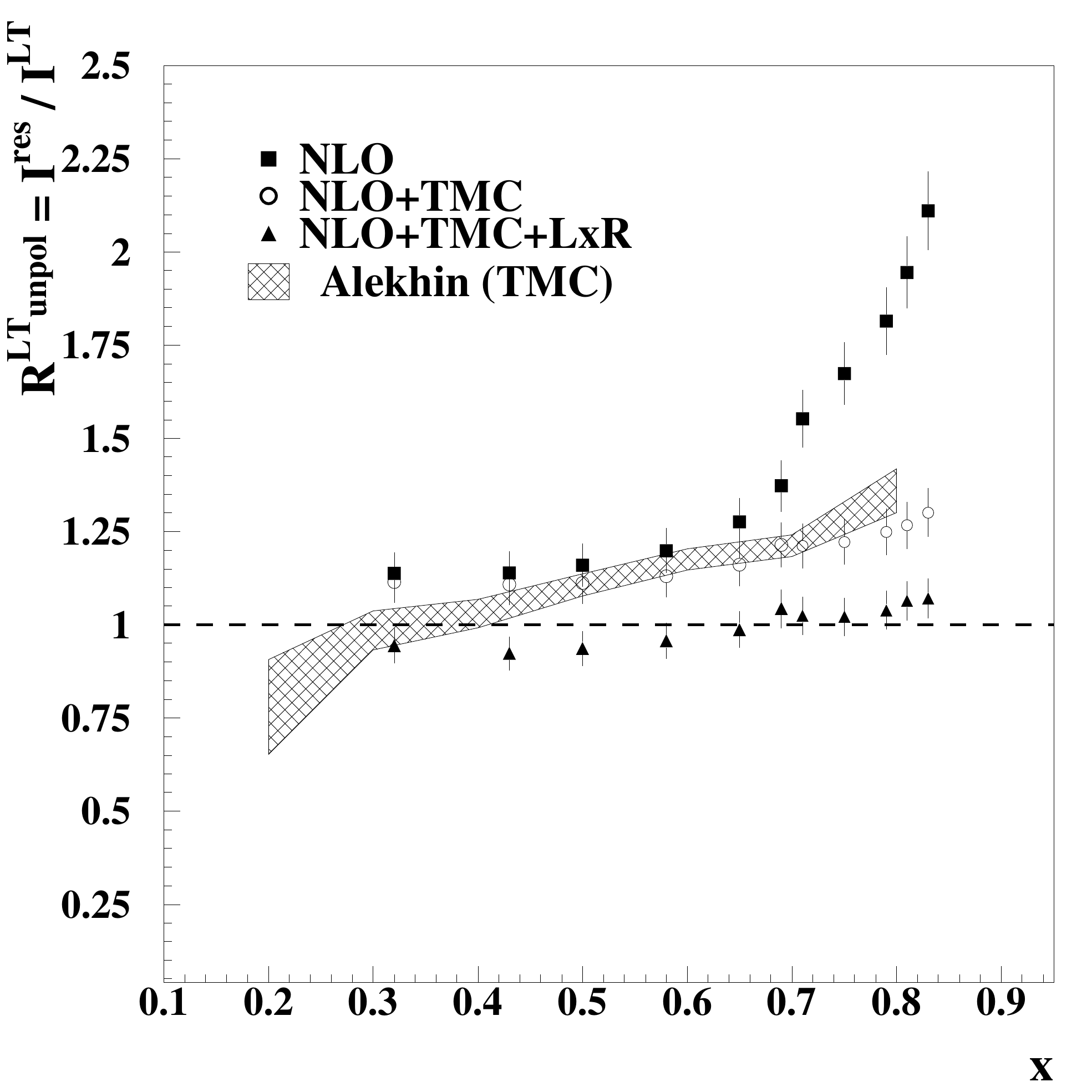}
  \caption{HT coefficients extracted in the resonance region according 
         to the procedure described in the text.  
         HT extracted with only the NLO calculation (squares); the 
         effect subtracting TMC (open circles); the effect of subtracting both 
         TMC and LxR (triangles). Shown for comparison are the values obtained from 
         the coefficient $H$ obtained in Ref. \protect\cite{Ale1} using DIS data and 
         including the effect of TMC.  Adapted from Ref. \protect\cite{BFL}}
\end{figure}
Since we extend our $x$-dependent analysis to the resonance region
we consider the following integrated quantities
\begin{equation}
\label{Iexp}
I^{\mathrm{res}}(\langle x \rangle, Q^2) = \int^{x_{\mathrm{max}}}_{x_{\mathrm{min}}} 
F_2^{\mathrm{res}}(x,Q^2) \; dx
\end{equation}
where $F_2^{\mathrm{res}}$ is evaluated using the experimental data  
in the resonance region. 
For each $Q^2$ value:
$x_{\mathrm{min}}=Q^2/(Q^2+W_{\mathrm{max}}^2-M^2)$, and 
$x_{\mathrm{max}}=Q^2/(Q^2+W_{\mathrm{min}}^2-M^2)$, where 
$W_{\mathrm{min}}$  and $W_{\mathrm{max}}$ delimit the resonance region,
and $\langle x \rangle$ is the average 
value of $x$ for each kinematics. 
This procedure replaces a strict
point by point in $x$, analysis.
\begin{figure}
\label{fig2}
  \includegraphics[height=.45\textheight]{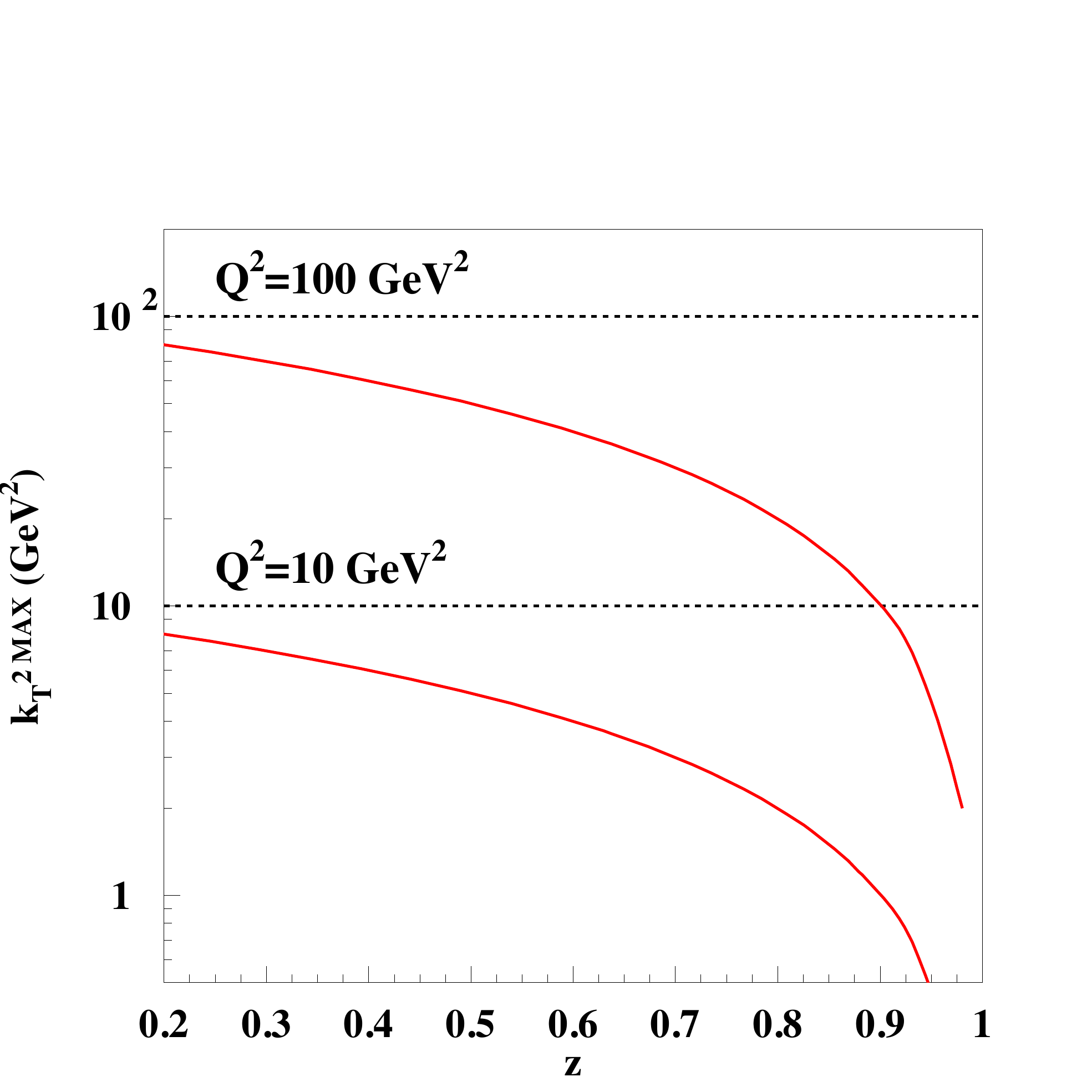}
  \caption{Phase space in the evolution of the NS component. The dotted lines are for $k_{T, \, MAX}^2 = Q^2$, for $Q^2 = 10, 100$ GeV$^2$. 
  The full lines represent the upper limit $k_{T, \, MAX}^2 = Q^2(1-z)/z$ for $Q^2=10, 100$ GeV$^2$. In this case, one can see a clear reduction 
  of the allowed $k_T$ at large $z$.}
  \end{figure}

Typical results from the analysis outlined above are plotted in Figure 1 where we show the HT coefficient  defined from Eq.(\ref{t-exp}) as
\[ R^{LT} \equiv C(x) = Q^2 \left[F_{2}(x,Q^2)/F_{2}^{LT}(x,Q^2) -1 \right]
  \]
The error in the figure is from the experimental data. No theoretical uncertainty was included.  However, our results clearly show that the combined effects of 
TMCs and LxR substantially reduce $C(x)$. We take this as illustrative of the accomplishments one can expect from the analysis we suggest in this contribution.
Essential features that emerge are the interplay between the values of $\alpha_S(M_Z^2)$ and the HTs, the relevance of TMCs, and, most importantly,  the need to
define $\alpha_S$ in the infrared region.  All of these features can affect the central values of the HTs reported in Fig. 1. 
\begin{figure}
\label{fig3}
  \includegraphics[height=.4\textheight]{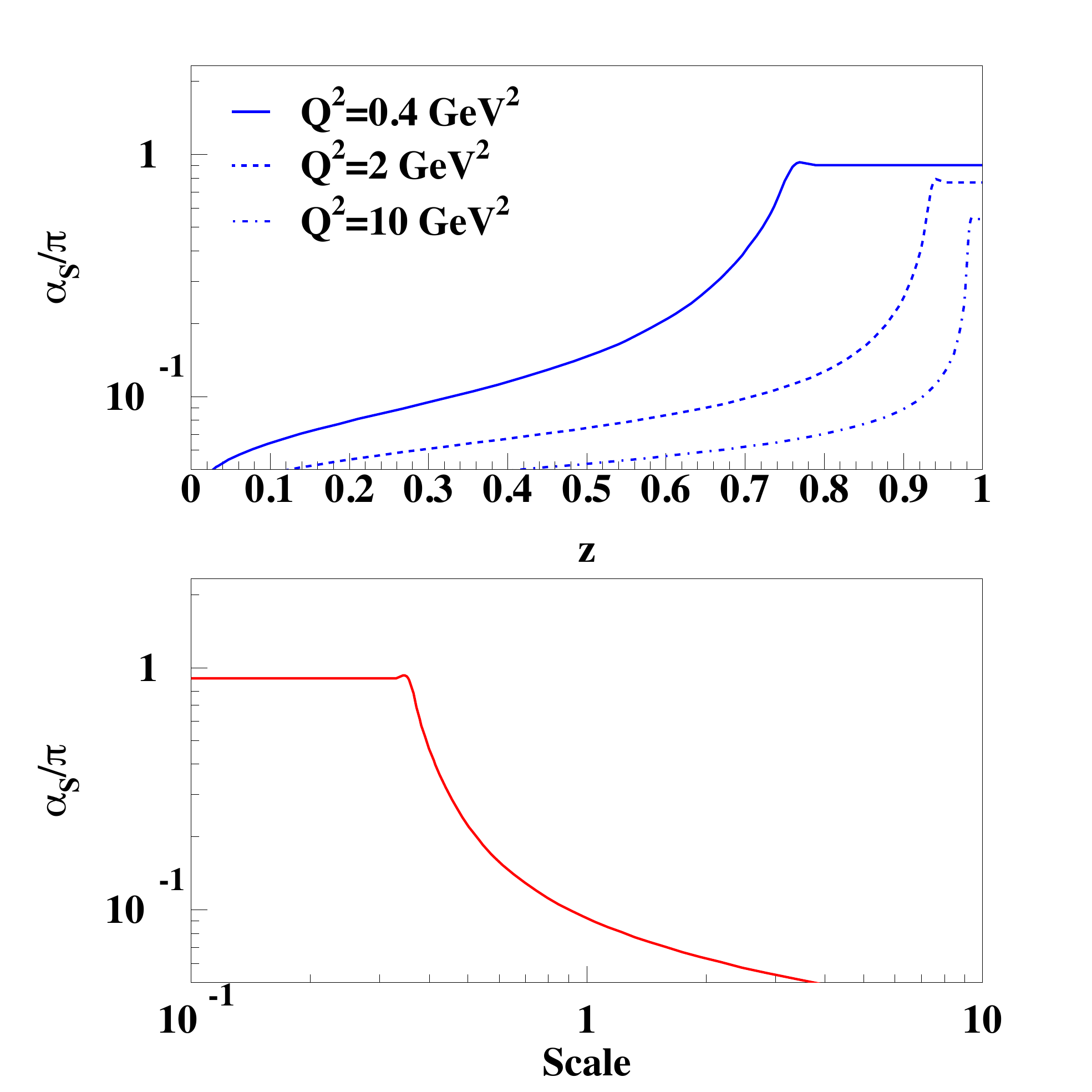}
  \includegraphics[height=.4\textheight]{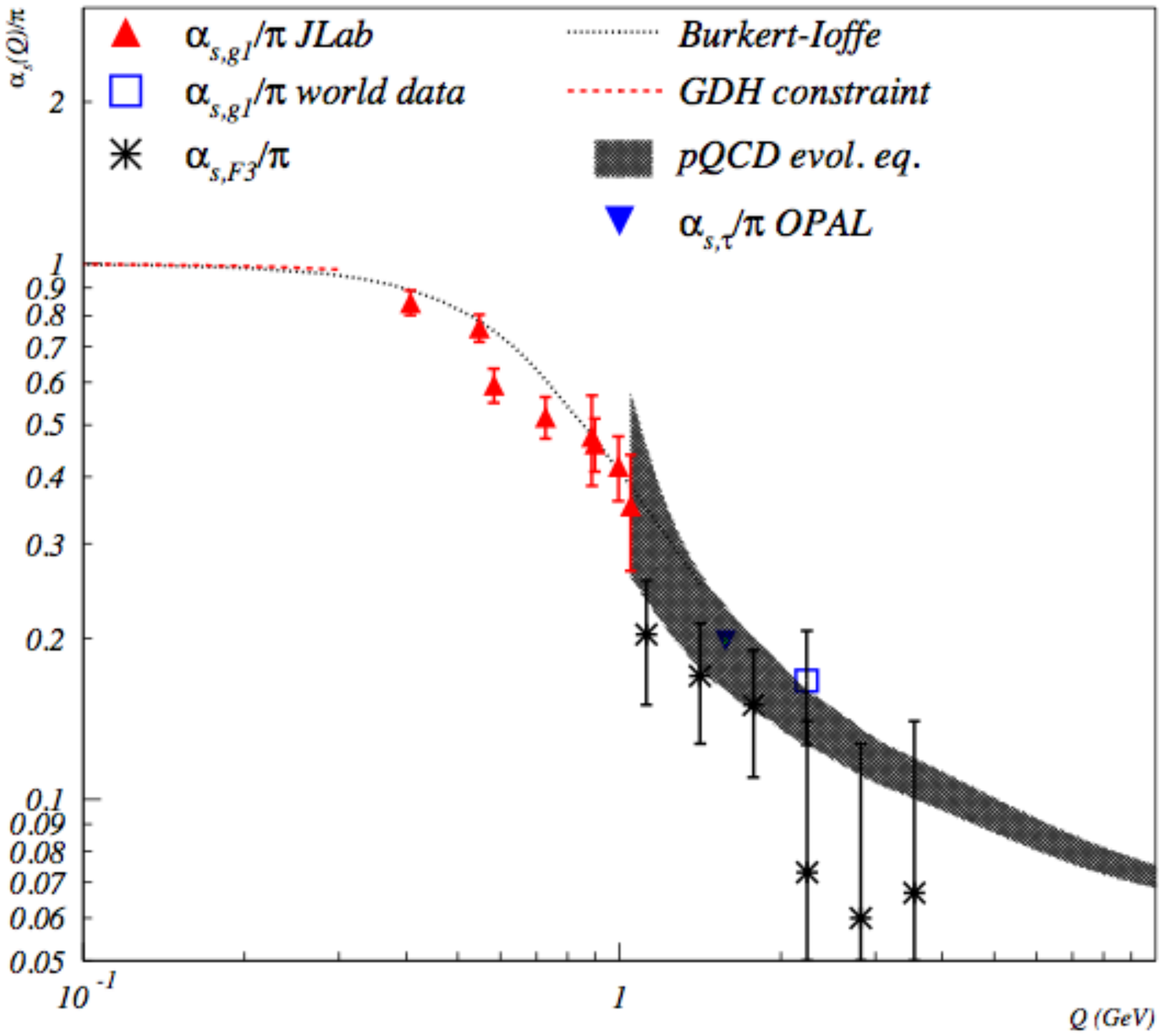}
  \caption{Left panel: $\alpha_S/\pi$ extracted from the analysis of the large $x$ data discussed in the text, and plotted
  vs. $z$, Eq.(\ref{lxr}) (upper panel), and vs. $\widetilde{W} = \sqrt{Q^2(1-z)/z}$ (lower panel). For comparison 
  we show the extraction from Ref. \protect\cite{ChenDeur} using Jefferson Lab data at $Q^2 = 0.7-1.1$ GeV$^2$.}
  \end{figure}
  
\section{Strong coupling constant at high $x$}
We now discuss in more detail the working of threshold resummation, and its possible impact 
on the analysis of $F_2$ at large $x$  \cite{Rob}. Starting from NLO, the coefficient, $C(z)$ in Eq.(\ref{lxr}) 
is dominated at large $x$ by terms proportional to $[\alpha_S(Q^2) \ln(1-z)]^n$ which need to be resummed
in the perturbative series. The physical origin of these terms is in the phase space for the contribution
of gluons emission to evolution, which become soft as $x\rightarrow 1$. A mismatch in the cancellation
with the virtual gluons contributions ensues.  
If, however, one carefully evaluates the kinematics for gluon emission at large $x$ 
within a quark-parton model view, one obtains \cite{BroLep}
\begin{eqnarray}
q(x,Q^2) & = & q(x,Q_o^2) + \int_{Q_o^2}^{\widetilde{W}^2} \frac{dk_\perp^2}{k_\perp^2}  
\frac{\alpha_S(k_\perp^2)}{2\pi}
\int_x^1 \frac{dz}{z} P_{qq}\left( \frac{x}{z},
\alpha_S \right) q(z,k_\perp^2),
\end{eqnarray}
where $Q^2_o$ is an arbitrary initial scale, and $\widetilde{W}^2= Q^2(1-z)/z$ is the maximum $k_\perp^2$ in the virtual photon-quark
center of mass system, appearing in the ladder graphs that define the leading log result. The resulting phase space is shown for
different $Q^2$ values in Fig. \ref{fig2}. 
The reduction of the allowed $k_\perp^2$ results  in a simultaneous shift in the argument of $\alpha_S \rightarrow \alpha_S(Q^2/(1-z)/z)$,
and a cancellation of the $\alpha_S(Q^2) \ln(1-z)$ divergence in the NLO coefficient function.  
As a consequence of rescaling the argument of $\alpha_S$ one has to consider its continuation into the infrared region \cite{Rob1,BFL}. 
The left panels of Fig. \ref{fig3} display the results for $\alpha_S$ used in our analysis for different values of $Q^2$. We also show, on the right, the extracted
value of the effective $\alpha_S$ from the GDH sum rule. We therefore suggest large $x$ evolution in DIS as yet another way of defining  an effective coupling constant at low values of the 
scale. A more quantitative analysis to relate different types of measurements, and to study in depth the possible process dependence of $\alpha_S$ is in 
progress \cite{ChenDeurLiu}.   

\section{Conclusions}
In conclusion, we believe there is a much richer structure to the scale dependence of the nucleon's distribution functions
that persists
behind the apparent cancellation among higher twist terms.
We started uncovering this structure in the initial work of Refs.\cite{Kep1,BFL}. 
Our analysis opens up the possibility of extracting 
the effective strong coupling at low scale, and to connect it to extractions from 
different processes, namely Refs.\cite{ChenDeur,Deur1} and Ref.\cite{Courtoy}.
While on one side this points at the fact that PQCD provides an essential framework for understanding the working of duality, 
a possible connection with the model of Refs.\cite{CloIsg,CloMel,Jes,CloZhao} is envisaged through the definition of effective 
charge \cite{ChenDeurLiu}.

\section*{Acknowledgments}

This work is supported by  the U.S. Department
of Energy grant no. DE-FG02-01ER41200. Many discussions with J.P. Chen, A. Courtoy, A. Deur and V. Vento are
gratefully acknowledged.


\end{document}